# HST and MERLIN Observations of the Jet in 3C273[1]

J.N. Bahcall and S. Kirhakos

Institute for Advanced Study, School of Natural Sciences, Princeton, NJ 08540

D.P. Schneider

Department of Astronomy and Astrophysics, The Pennsylvania State University, University Park, PA 16802

R.J. Davis[2] and T.W.B. Muxlow and S.T. Garrington and R.G. Conway

The University of Manchester, Nuffield Radio Astronomy Laboratories, Jodrell Bank, Cheshire, UK

and

S.C. Unwin

Caltech 105-24, Pasadena, CA 91125

## ABSTRACT

We present red and blue images of the jet of the quasar 3C273 obtained with the WFPC2 on the Hubble Space Telescope as well as a new radio map made with the MERLIN array. The images are of significantly better quality than that of previous data. The two maps are aligned to an accuracy of $0.020''$; this accuracy is achieved because both the quasar and the jet are contained in both the radio and optical images. The start of the optical jet is marked by an elongated knot which appears identical at radio and optical wavelengths. Other knots in the optical jet correspond to narrow oblique features within the radio outline. The total width of the smooth emission in the optical jet is $0.7''$; the FWHM of the optical knots is $0.3''$. The knots may trace the current location of a narrow, perhaps helical jet lying within the outlines of the older radio cocoon.

*Subject headings:* Quasars: general; quasars:individual: 3C273

astro-ph/9509028  06 Sep 1995

## 1. Introduction

The jet near 3C 273 is the only extended, linear feature observed at both radio and optical wavelengths for a luminous quasar. Ground-based investigations (Röser & Meisenheimer 1991, Conway et al. 1993, Conway & Davis 1994) have suggested that the optical emission, like the radio, is produced by the synchrotron process.

Previous observations with the Hubble Space Telescope (Thomson, Mackay, & Wright 1993) showed that the optical jet is narrow and consists of discrete knots, but these data suffer from the effects of spherical aberration in the HST mirror. We present here a new optical image obtained with the refurbished HST, and compare it with a new radio map at 18 cm wavelength made with the improved MERLIN array. Both images are of significantly better quality than before, preserving high surface brightness sensitivity at $0.1''$ resolution. Since the present HST data were obtained with the much-improved optical characteristics of the post-repair mission, we can carry out quantitative photometry on the new HST images. Furthermore, since the quasar and jet both appear in the same field in the new data, the images may be accurately aligned, which makes possible a detailed comparison of the radio and optical features of the jet.

As is well known, the innermost portion of the jet of 3C273 contains features showing 'superluminal' motion (Unwin et al. 1985) which must be moving at relativistic velocities at an angle close to the line of sight. Unlike the majority of quasars, 3C273 appears one-sided with no sign of a counter-jet either on parsec or kiloparsec scales (Davis, Muxlow, & Conway 1985). The observed superluminal motion implies that Doppler beaming could hide a counter-jet on parsec scales. However, it is not clear whether the lack of an extended counter-jet can be explained by Doppler beaming alone, or requires that the system is intrinsically asymmetric. The high degree of radio polarization of the end of the jet is hard to explain if seen end-on (Conway & Davis 1994), suggesting that the system is not a simple axisymmetric flow. The new observations at radio and optical wavelengths presented here also suggest a complex system of flow.

We assume in this paper that $H_0 = 75 \,\mathrm{km\,sec^{-1}\,Mpc^{-1}}$ and $q_0 = 0.5$; at the distance of 3C273 ($z = 0.158$), $1''$ corresponds to 2.37 kpc.

## 2. Observations

The quasar 3C273 was observed on 4-5 June, 1994 with the Wide Field/Planetary Camera-2 (WFPC2) of the Hubble Space Telescope. Three separate exposures (1100 s, 600 s, and 100 s) were obtained through the F606W filter which has a bandpass similar to $V$ ($\lambda_c = 5935$ Å, $\Delta\lambda = 1497$ Å). Additional exposures (900 s and 200 s) were obtained with the F450W filter (similar to $B$; $\lambda_c = 4519$ Å, $\Delta\lambda = 956$ Å). The properties of the WFPC-2, based on the first year of operation, are given in Burrows 1995. All magnitudes in this paper are based on the photometric system described in Burrows 1995, where objects with a constant energy distribution in $f_\nu$ have neutral color, and the zero point is set so that the magnitude at $\approx 5500$ Å is similar to the standard $V$ magnitude.

The initial data processing (bias frame removal and flat-field calibration) was performed using standard software at the Space Telescope Science Institute; the STScI photometric calibration (see Burrows 1995) was verified by comparing galaxy magnitudes with ground-based values. All frames of a given filter were aligned to much better than a pixel, so it was easy to identify and eliminate cosmic ray events. The image scale in the center of the detector is 0.0.0996″ pixel$^{-1}$; this large image scale (relative to the diffraction limit) reduces the effective resolution to approximately this value, but the resulting large field permits both the quasar and the entire jet to be imaged in one exposure with a single detector.

The 1400 s F606W image is displayed in Figures 1 and 2a. The quasar is heavily saturated and is offset by approximately 5″ from the field center. The main jet consists of a number of high surface brightness knots extending from 13″ to 21″ from the quasar. The sky brightness in this image is 22.1 mag arcsec$^{-2}$ and the limiting magnitude for point sources is $\sim$25.0 mag. The values for the sky brightness and limiting point source magnitudes in the F450W image are 22.7 mag arcsec$^{-2}$ and $\sim$ 24.0 mag, respectively.

MERLIN observations were made in May 1993 at 1658 MHz (wavelength 18 cm), with a bandwidth of 30 MHz, of which a single 1 MHz channel was selected for mapping. The MERLIN array included the 76-m Lowell telescope and the new 32-m telescope at Cambridge, giving a maximum baseline of 217 km. The visibility amplitudes were calibrated relative to 3C286 (assumed flux density 13.65 Jy, Baars et al. 1977), using an unresolved point source as an intermediary calibration. After several cycles of self-calibration to correct phase and amplitude variations, the radio image was obtained by deconvolution using a combination of CLEAN and MEM. The resolution corresponds to a beamwidth of 0.18″ × 0.14″ (FWHM) at position angle 27°. The limiting sensitivity of the radio image is shown by the lowest plotted contour in Fig. 1 which is 1.5 mJy/beam, approximately three times the rms noise.

## 3. Results

The HST image reveals that the optical jet consists of a series of knots (or components), linked by fainter emission, all confined to a narrow tube. The knots are labeled in Fig. 2a, except for the knot closest to the quasar. The first knot (hereafter, F) is not shown in Fig. 2a, but it is coincident with the radio component closest to the quasar (separation of 5.5″) that appears in Fig. 1. The components A1, B1, and C3 are elongated along the main axis of the jet; knots B2, C1, C2, and H3 lie oblique to the jet axis.

The optical surface brightness of the smooth emission of the jet between knots A1, A2 and B1 is $\sim 23.0$ mag arcsec$^{-2}$ in the F606W image. It is more difficult to measure the surface brightness between other knots, since they are so closely crowded together. However, within the uncertainties ($\sim 0.3$ mag), the surface brightness of the diffuse, or smooth, component is essentially constant.

The surface brightness of the jet perpendicular to its extent is flat-topped. It is approximately constant within the region of emission and falls to zero over a distance smaller than one pixel. We have measured the total width, $w_{\rm tot}$, at five different locations in the smooth region and find a width $w_{\rm tot} = 0.7'' \pm 0.1''$. The surface brightness in the knots is more peaked; for the knots, we find a FWHM of $w_{\rm FWHM} = 0.3'' \pm 0.1''$. For the adopted cosmological parameters, $w_{\rm tot} = 1.7$ kpc and $w_{\rm FWHM} = 0.7$ kpc.

Aperture magnitudes from the HST data for each of the individual knots are given in Table 1, which lists the jet component, its angular distance from the center of light of the quasar, the diameter of the aperture, the magnitude in the F606W band, and the color between bands F450W and F606W. The principal source of error in these measurements is the determination of the background level. The uncertainties in the values in Table 1 are estimated to be a few tenths of a magnitude. We cannot establish if components A3, In1, In2 and Ex1 are physically associated with the jet; component Ex1 is significantly redder than the other jet knots and might be a background galaxy.

The radio image is shown in Fig 2b: we confirm that the radio outline is both longer and wider than the optical (Röser & Meisenheimer 1991). The improvement shown in Fig 2b over previous radio images in both resolution and image quality is due partly to improved data processing methods, and partly to enhancements of the MERLIN array, such as the use of cooled receivers, and the inclusion of the new 32-m telescope in the array.

The radio image includes a portion of the jet located 0.5″ from the quasar, which is an extension of the parsec-scale jet seen by VLBI (Unwin et al. 1985). A more detailed study of the MERLIN and VLBI data in combination is being prepared, and will appear elsewhere together with a more detailed theoretical interpretation.

The optical and radio images were registered by aligning the position of the quasar. The centroid position of the radio image can be determined to within a few milliarcsec (mas). The optical position of the quasar was determined to within 10 mas (a tenth of a pixel) by locating the crossing of the diffraction spikes; this is similar to the size of the residuals that remain after representing the geometric distortion with a ten-term polynomial (Burrows 1995). The relative positioning in the optical of the quasar and the jet was more direct in the present work than in the previous discussion (see Thompson et al. 1993), because the quasar and the jet appear on the same WFPC2 image. We expect that the two data sets should be aligned to 20 mas. No features in the jet were used in the registration process. An overlay of the radio and optical images is shown in Fig. 1, and a portion, restricted to the jet and excluding the quasar, is shown in Fig. 3.

The radio jet has a constant half-power width in the radio of $1.0'' \pm 0.1''$ at separations of $13''$ to $20''$ from the quasar (cf. Conway et al. 1993). The apparent change in the radio width is an artifact caused by the change in radio brightness along the length of the jet. Previous maps at lower resolution have traced the jet all the way from the quasar and revealed a diffuse lobe to the south of the jet (Davis, Muxlow, & Conway 1985). In the present map, these features are not prominent, due to resolution by the radio interferometer, but considerable structure may be seen within the jet. In particular, there are a number of oblique features about $0.5''$ long and separated by $1.4''$, which are inclined at about 45 degrees to the axis of the jet. These can be seen both in the weak part of the jet in Fig. 2(b) and towards the bright head of the jet in Fig. 3.

## 4. Discussion

Direct comparison of the radio and optical images (Fig. 3) shows that the oblique radio features all coincide with optical knots, though there are some optical features without radio counterparts. The two extensions to the west of the optical jet (Lelièvre et al. 1984, Röser & Meisenheimer 1991, Thomson, Mackay, & Wright 1993) have no detectable radio emission. Conversely, the bright radio head is only barely detected optically.

Assuming that the optical emission is indeed incoherent synchrotron, the lifetime of the radiating electrons is 1000 times shorter at optical wavelengths than at radio wavelengths; if the magnetic field strength is 30 nT (Conway et al. 1981) the optical lifetime is only 1000 yr. Assuming a migration speed of $\leq 0.1c$, the optical electrons must be excited within 100 light-years of their present position, which corresponds to an angular scale of only 12 milliarcsec. Thus, the optical emission traces the present location of the active jet, while the radio emission provides information on its history during the past 1 Myr. This difference in age makes it puzzling that the oblique radio features should correspond so closely to the optical knots in angular size; a wavelength-independent model must be found to explain the enhancements in brightness along the jet.

We suggest that what has hitherto been called the 'radio jet' is actually two components superposed, firstly the fast-moving jet proper, shown by the coincident oblique radio and optical features, and secondly emission from a surrounding, slow-moving 'cocoon' (Begelman, Blandford, & Rees 1984). The 'lobe' seen at 408 and 151 MHz (Davis, Muxlow, & Conway 1985) may be an extension of this cocoon. Whether the cocoon material is stationary or is flowing backwards is not clear from our data, but in either case the associated Doppler effects in the cocoon will be negligible. The radio greyscale shown in Fig. 2(b) suggests that the oblique radio features coincident with the optical knots may be in the form of a helix. If the velocity is relativistic, the emission will appear brightest where the velocity vector is closest to the line of sight, the enhancement being independent of wavelength. This would explain the close correspondence between the optical and radio knots. The helical form may arise from a driven Kelvin-Helmholtz instability (Hardee, Cooper, & Clarke 1994). We note that some features of the model presented above are similar to the wind-model of Thomson et al. (1993).

In contrast to the knots, the bright radio head (component H2 of Flatters and Conway 1985) is very faint at optical wavelengths; the optical jet apparently stops 2″ short of the radio head. This means the enhancement of H2 relative to the remainder cannot simply be due to geometric factors, such as Doppler boosting, for if H2 was the terminal jet shock, we would expect it to be bright at both wavelengths. Perhaps the terminal shock is now at knot D, and H2 represents what has been termed a 'disconnection event' (Cox, Gull, & Scheuer 1991).

In conclusion, a detailed comparison of new high-resolution radio and optical images of 3C273 suggests that the true jet, located along the line of the optical knots, is accompanied by enhanced radio emission. This is superposed on more general radio emission from a surrounding cocoon, resulting from older material emerging from the jet. The variations of brightness along the jet itself may be governed by geometrical factors and suggest that the jet is in the form of a helix.

This work was supported in part by NASA contract NAG5-1618 (grant number GO-2424.01) from the Space Telescope Science Institute, which is operated by the Association of Universities for Research in Astronomy, Inc. under NASA contract NAS5-26555.

---

Figure Captions

Fig. 1.— Combined diagram of the radio and optical emission from the 3C273 system. The overexposed optical quasar image shows diffraction spikes, used to locate the optical center, together with a linear feature that arises due to charge leakage from the CCD. The radio contours are shown surrounding the optical emission and are at equal logarithmic intervals of a factor of two from 1.5 mJy beam$^{-1}$. The optical and radio resolutions are approximately $0.10''$ and $0.16''$, respectively.

Fig. 2.— Greyscale images of the jet in quasar 3C273 at optical and radio wavelengths. The upper panel (a) shows the optical image made at $\lambda 5935$ Å with WFPC2 on the Hubble Space Telescope. The resolution is limited by the CCD pixel size ($0.1''$). The lower panel (b) displays the radio image made with MERLIN at $\lambda 18$ cm. The radio resolution is $0.18 \times 0.14$ arcsec at PA $27°$.

Fig. 3.— An overlay of the optical (greyscale) and radio (contour) images of the jet. The radio contours are as in Fig. 1. Optical features (A–D) have been labeled using the nomenclature of Lelièvre et al. (1984) though these could now be subdivided, and the subcomponents of the radio head (h1–h3) have been labeled according to Flatters and Conway (1985).

Table 1. Optical photometry

| Component | D (″) | Aperture (″) | m606 (mag) | m(450)–m(606) |
|---|---|---|---|---|
| F | 5.5 | 1.0 | 24.4 | — |
| A1 | 13.0 | 1.0 | 22.5 | 0.3 |
| A2 | 14.2 | 0.8 | 23.4 | 0.3 |
| A3 | 12.7 | 0.6 | 24.8 | −0.2 |
| B1 | 15.2 | 0.6 | 23.3 | 0.2 |
| B2 | 15.8 | 0.6 | 23.6 | 0.2 |
| C1 | 16.8 | 1.0 | 22.8 | 0.6 |
| C2 | 17.7 | 1.0 | 22.8 | 0.2 |
| C3 | 19.0 | 1.0 | 22.7 | 0.5 |
| D | 19.9 | 0.8 | 22.6 | 0.6 |
| H3 | 20.4 | 1.0 | 22.3 | 0.4 |
| H2 | 21.3 | 0.6 | 24.2 | 0.5 |
| Ex1 | 22.5 | 1.0 | 24.1 | 0.8 |
| In1 | 12.2 | 0.6 | 24.2 | −0.1 |
| In2 | 12.4 | 0.6 | 24.0 | 0.0 |